\documentclass[10pt, aps, prd,
reprint, 
notitlepage, nofootinbib, longbibliography, showpacs, letter,
superscriptaddress,
]{revtex4-1}

\usepackage{pstricks}
\usepackage{amsmath,bm}
\usepackage{graphicx}
\usepackage{verbatim}
\usepackage[colorlinks=true,urlcolor=blue,linkcolor=red,citecolor=red]{hyperref}

\begin{document}

\title{Casimir energies of self-similar plate configurations}

\author{K. V. Shajesh}
\email{kvshajesh@gmail.com} \homepage{http://www.physics.siu.edu/~shajesh}
\affiliation{Department of Physics, Southern Illinois University--Carbondale,
Carbondale, Illinois 62901, USA}
\affiliation{Department of Energy and Process Engineering,
Norwegian University of Science and Technology, N-7491 Trondheim, Norway}

\author{Iver Brevik}
\email{iver.h.brevik@ntnu.no}
\homepage{http://folk.ntnu.no/iverhb}
\affiliation{Department of Energy and Process Engineering,
Norwegian University of Science and Technology, N-7491 Trondheim, Norway}

\author{In\'{e}s Cavero-Pel\'{a}ez}
\email{cavero@unizar.es} \homepage{http://cud.unizar.es/cavero}
\affiliation{Centro Universitario de la Defensa (CUD),
Zaragoza 50090, Spain}

\author{Prachi Parashar}
\email{prachi@nhn.ou.edu}
\affiliation{Department of Physics, Southern Illinois University--Carbondale,
Carbondale, Illinois 62901, USA}
\affiliation{Department of Energy and Process Engineering,
Norwegian University of Science and Technology, N-7491 Trondheim, Norway}

\date{\today}

\begin{abstract}
We construct various self-similar configurations using parallel
$\delta$-function plates and show that it is possible to evaluate 
the Casimir interaction energy of these configurations using the 
idea of self-similarity alone. We restrict our analysis to 
interactions mediated by a scalar field, but the extension to 
the electromagnetic field is immediate. Our work unveils an easy and 
powerful method that can be easily employed to calculate the Casimir 
energies of a class of self-similar configurations. As a highlight, 
in an example, we determine the Casimir interaction energy of a 
stack of parallel plates constructed by positioning $\delta$-function 
plates at the points constituting the Cantor set, a prototype of a 
fractal. This, to our knowledge, is the first time that the Casimir 
energy of a fractal configuration has been reported. Remarkably, 
the Casimir energy of some of the configurations we consider turn 
out to be positive, and a few even have zero Casimir energy. For 
the case of positive Casimir energy that is monotonically decreasing 
as the stacking parameter increases the interpretation is that the 
pressure of vacuum tends to inflate the infinite stack of plates. We
further support our results, derived using the idea of self-similarity 
alone, by rederiving them using the Green's function formalism. These
expositions gives us insight into the connections between the 
regularization methods used in quantum field theories and 
regularized sums of divergent series in number theory.
\end{abstract}

\maketitle

\section{Introduction}

Physical phenomena associated with the interaction energy between
two bodies, arising as a direct manifestation of the quantum
fluctuations in the field mediating the interactions, is broadly 
termed the Casimir energy. The Casimir force between two parallel
conducting plates associated with this interaction energy was 
first theoretically predicted by Casimir in Ref.~\cite{Casimir:1948dh}.
In this article, for simplicity in the mathematical analysis,
we consider the interactions to be mediated by a scalar field. 
Nonetheless, many of the physical interpretations and intuition 
we have amassed for the electromagnetic field can often be extended 
to the scalar model,
especially for the case of perfect conductors because one of the modes
for the electromagnetic case can be represented by a scalar field
satisfying Dirichlet boundary conditions. Since the original 
calculation by Casimir, the Casimir energies of special geometries 
like parallelepipeds~\cite{Lukosz:1971,Ambjorn:1981xw,Ambjorn:1981xv},
spheres~\cite{Boyer1968pc,Milton:1978sf}, and 
cylinders~\cite{DeRaad1981229,CaveroPelaez:2004xp}
has been reported both for the scalar field and for the electromagnetic field.
More recently, using the multiple scattering 
formulation~\cite{Balian1977300,Balian1978165}
the single-body contributions were generically separated from the 
total energy~\cite{Kenneth2006fc,Emig:2007me,Milton:2007wz}, and
it has become possible to compute Casimir
energies for arbitrary shaped disjoint objects.
An extension of the theory so as to include  dynamical Casimir effects 
leads to fundamental quantum mechanical phenomena such as 
Casimir friction, c.f., for instance, 
the recent review of Ref. \cite{symmetry16}.

A generalization of these ideas to more than two bodies was given in 
Refs.~\cite{Schaden:2010wv,Shajesh:2011ef,Shajesh:2011rv}, but
explicit solutions for the Green's functions were reported only for 
configurations with three bodies. Here, in Sec.\,\ref{sec-Gfuns},
we find solutions to the Green's function for four bodies, and then, we
go further and express the solution to the Green's function for $N$ bodies 
as a recursion relation in terms of the Green's functions for $(N-2)$ bodies.
This procedure then lets us extend our solutions for the Green's functions
for an infinite sequence of objects by taking the limit $N\to\infty$.

In Sec.\,\ref{sec-CaseI},
we use the solution for the Green's function for an infinite sequence
of objects to calculate the Casimir energy of self-similar configurations.
In particular, we calculate the Casimir energy of parallel
$\delta$-function plates satisfying Dirichlet boundary conditions
that are positioned in various patterns to construct simple
self-similar configurations. The Casimir energy of these infinite
sequence of plates comes to be positive, negative, or zero,
suggesting that the pressure of vacuum tends to inflate, deflate,
or balance the infinite stack of plates.
These results are obtained by regularizing sums for divergent series,
which on its own might not be convincing.
But, the highlight of this article, is that we are able to 
derive all of the above results using the idea of self-similarity 
alone in a self-contained manner. We begin our discussion 
in Sec.~\ref{sec-self-sim} by presenting these derivations 
using the idea of self-similarity,
which we again point out are completely independent of the derivation
using Green's function formalism that we apply later 
in Secs.\,\ref{sec-Gfuns} and \ref{sec-CaseI} to further support our claims.
In Sec.\,\ref{sec-rel-string}, we present an analogy between our present
study and the theory of the piecewise uniform string.
In Sec.\,\ref{sec-con-out}, we present few concluding remarks and an outlook.

\section{Casimir interaction energies for self-similar configurations}
\label{sec-self-sim}

A self-similar set contains the set itself as a subset, or more
generally, there exists a one-to-one mapping between the elements
of the set and a subset of the set.
The property of self-similarity is illustrated well when it is used 
to sum a series. Consider an infinite sum 
\begin{equation}
x=1+\frac{1}{2} +\frac{1}{4} +\frac{1}{8} + \ldots.
\end{equation}
Using the idea of self-similarity we can identify the following relation 
for the sum,
\begin{equation}
x = 1 + \frac{1}{2} x,
\label{x=1+12x}
\end{equation}
which immediately leads to the conclusion that the sum of the series is $x=2$.
We can extend this idea of self-similarity to `sum' a divergent series too.
For example, for the divergent sum 
\begin{equation}
x=1+2+4+8+\ldots,
\end{equation}
using the idea of self-similarity, we can identify the relation 
\begin{equation}
x=1+2x,
\end{equation}
which assigns the value $x=-1$ to the above divergent sum
and is interpreted as the `sum' of the divergent series.
Even though values assigned to divergent series in this manner
are now well accepted as a regularized sum, the perplexities 
associated with these manipulations 
in the spirit of Ref.~\cite{Hardy-b1109376} still linger on.
Here, we construct self-similar configurations of parallel plates,
and using the idea of self-similarity along the lines of the illustrations
above, we derive the Casimir interaction energies for these configurations.

We construct a planar configuration consisting of an infinite sequence
of parallel $\delta$-function plates. 
By $\delta$-function plates we mean infinitely thin plates that are
mathematically described by Dirac $\delta$-functions.
The position of the plates are given by the sequence
\begin{equation} 
a_1, a_2, a_3, \ldots, 
\end{equation}
the `strength' of the plates are given by the sequence
\begin{equation} 
\lambda_1, \lambda_2, \lambda_3, \ldots,
\end{equation}
and their interactions are mediated through a scalar quantum field, 
with the plates described by the potentials
\begin{equation}
V_i({\bf x})=\lambda_i \delta(z-a_i).
\end{equation}
When the dynamics of the plates is neglected (valid when the 
masses of the plates are large), the vacuum to vacuum 
transitions induced by the quantum fluctuations of the scalar field
leads to energy and momentum densities, which are given in terms 
of the energy-momentum tensor and the associated Green's function
for the scalar field. The total energy, obtained by integrating the 
energy density over all space, is termed the vacuum energy or the Casimir
energy or the zero point energy. 
Here, we discuss the Casimir energy of an infinite sequence of
parallel $\delta$-function plates. 

\begin{figure}
\begin{center}
\includegraphics{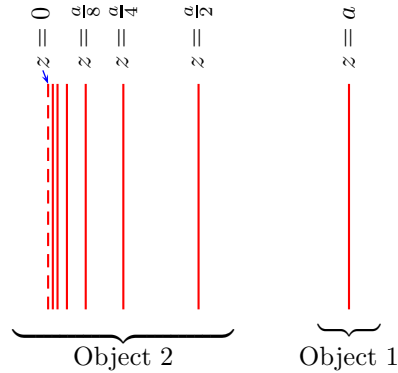}
\caption{A geometric sequence of parallel plates. 
The position of the plates is given by the sequence
$z=a, \frac{a}{2}, \frac{a}{4}, \frac{a}{8}, \ldots$.
The first seven plates of the infinite sequence have been shown.
The dashed line to the left is the limit of this sequence of plates. 
}
\label{fig-a2i-plates-12}
\end{center}
\end{figure}%

It is well known, for example, see 
Ref.\,\cite{Balian1977300,Balian1978165,Ravndal:2000kn,Bulgac:2001ef,Bulgac:2006sp,Kenneth2006fc,Shajesh:2011ef},
that the total energy per unit area, ${\cal E}=\text{Energy/Area}$, 
for two parallel plates separated by distance $a$, can be decomposed
in terms of the respective one-body energies as 
\begin{equation}
{\cal E} = {\cal E}_0 + \Delta {\cal E}_1 
+ \Delta {\cal E}_2 + \Delta {\cal E}(a),
\label{en-mbbu}
\end{equation}
where ${\cal E}_0$ is the energy of the vacuum in the absence 
of the two objects, $\Delta {\cal E}_i= {\cal E}_i- {\cal E}_0$, 
$i=1,2$, are the one-body energies
associated to the individual objects, and $\Delta {\cal E}(a)$ is the 
interaction energy per unit area of the plates. In general, the one-body
energies and the bulk energy ${\cal E}_0$ diverge, and the Casimir interaction
energy per unit area $\Delta {\cal E}(a)$ is finite 
and is distinctly isolated by its dependence on the distance $a$, 
a signature of the interaction between the two plates.
The Casimir interaction energy between two plates, mediated through a scalar
field satisfying Dirichlet boundary conditions on the plates is given by,
for example see Refs.~\cite{Elizalde1991ci,Milton:2001zpe},
\begin{equation}
\Delta {\cal E}_{12}(a)=-\frac{\pi^2}{1440 a^3},
\label{Ce-twop}
\end{equation}
which is exactly half of the Casimir interaction energy for two 
perfectly conducting plates mediated through electromagnetic fields.
We are primarily interested in the interaction energy term in
Eq.~(\ref{en-mbbu}), which for multi-object configurations
will get many-body contributions. We do not bother to separate
this interaction energy into two-body, three-body, etc., like 
in Ref.\,\cite{Shajesh:2011ef}, and evaluate the total
Casimir interaction energy. After all the one-body contributions
has been subtracted, in addition to the bulk energy ${\cal E}_0$,
the remaining Casimir interaction energy is in general finite,
unless any two plates come into contact.

Let us consider an infinite sequence of plates
placed at the following positions:
\begin{equation}
z=a, \frac{a}{2}, \frac{a}{4}, \frac{a}{8}, \ldots,
\label{zpos-gpd}
\end{equation}
see FIG.~\ref{fig-a2i-plates-12}, such that the distances between the
plates successively decrease by a factor of two.
Let us analyze the energy break up of this infinite sequence of plates
by interpreting the single plate at $z=a$ as Object 1
and the rest of the plates to constitute Object 2.
Using the decomposition of energy in Eq.~(\ref{en-mbbu}), we can write
\begin{eqnarray}
\left( {\cal E}_0 
+\sum_{i=1}^\infty \Delta {\cal E}_i +\Delta {\cal E}(a) \right)
= {\cal E}_0 + \Delta {\cal E}_1
\hspace{20mm} \nonumber \\ \hspace{1mm}
+ \left( \sum_{i=2}^\infty \Delta {\cal E}_i + \Delta {\cal E}(a/2) \right)
+\Delta {\cal E}_{12}(a), \hspace{5mm}
\label{finCeay}
\end{eqnarray}
where we have isolated the single-body contributions to the energy
explicitly. The single-body contributions, in this manner, cancel out
in Eq.~(\ref{finCeay}) to give
\begin{equation}
\Delta{\cal E}(a) = \Delta{\cal E}(a/2) + \Delta{\cal E}_{12}(a),
\label{gp-ce-aa212}
\end{equation}
which requires some elaboration because we have used the idea of
self-similarity in writing Eq.~(\ref{gp-ce-aa212}).
The interaction energy of the complete stack of plates 
in FIG.~\ref{fig-a2i-plates-12} is on the left of Eq.~(\ref{gp-ce-aa212}).
The first term on the right of Eq.~(\ref{gp-ce-aa212}) 
is the interaction energy of the plates constituting Object 2
in FIG.~\ref{fig-a2i-plates-12}. And, the second term  
on the right of Eq.~(\ref{gp-ce-aa212}) is the interaction energy
between Object 2 and Object 1.
The idea of self-similarity has been used to note that the energy of
Object 2 is equal to the energy of the complete stack 
evaluated for a rescaled parameter, here $a/2$.
The interaction energy is a function of $a$ alone (for Dirichlet plates)
because that is the only parameter in the problem,
and on dimensional grounds, we can argue that 
\begin{equation}
\Delta {\cal E}(a/2)=2^3\,\Delta {\cal E}(a).
\label{scaArcs}
\end{equation}
Using the scaling argument of Eq.~(\ref{scaArcs}) in Eq.~(\ref{gp-ce-aa212}),
we identify the relation involving the Casimir interaction
energy of the infinite sequence of plates in FIG.~\ref{fig-a2i-plates-12},
\begin{equation}
\Delta {\cal E}(a) = 8\,\Delta{\cal E}(a) +\Delta{\cal E}_{12}(a),
\label{1ea=8ea+12ea}
\end{equation}
which is the analog of the relation for infinite series in Eq.~(\ref{x=1+12x}),
here for the Casimir interaction energies.

\begin{figure}
\begin{center}
\includegraphics{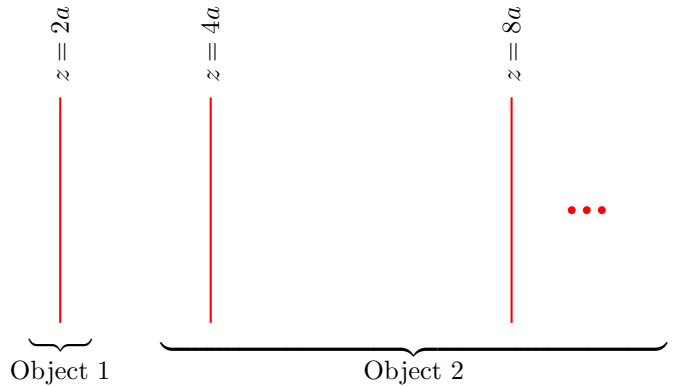}
\caption{A geometric sequence of parallel plates.
The position of the plates is given by the sequence
$z=2a, 4a, 8a, 16a, \ldots$.
}
\label{fig-a2i-plates-in}
\end{center}
\end{figure}%

The relation in Eq.~(\ref{1ea=8ea+12ea}) allows us to evaluate
$\Delta{\cal E}(a)$ in terms of the interaction energy between Object 1
and Object 2 given by $\Delta{\cal E}_{12}(a)$.
In general it is a difficult task to evaluate the interaction energy
$\Delta{\cal E}_{12}(a)$. But,
if each of the individual plates in the stack satisfy Dirichlet 
boundary conditions, which are called Dirichlet plates,
there is considerable simplicity in the analysis because 
a Dirichlet plate physically disconnects the two spaces across it.
For example, explicit decomposition of the total energy in terms of 
single-body, two-body, and three-body energies, and how they conspire 
such that the Casimir interaction energy is given completely
in terms of interaction of two Dirichlet plates was described in detail
in Ref.\,\cite{Shajesh:2011ef}. As a consequence,
each Dirichlet plate can only interact with its closest neighbor
on the left and on the right. Thus,
the interaction energy ${\cal E}_{12}(a)$
between the two bodies in FIG.~\ref{fig-a2i-plates-12}
is given by the Casimir interaction energy of two Dirichlet 
plates of Eq.~(\ref{Ce-twop}), separated in this case by distance $a/2$,
which is the distance between the plates at $z=a$ and $z=a/2$
in FIG.~\ref{fig-a2i-plates-12}.
Thus, we have
\begin{equation}
\Delta{\cal E}(a) = 8\,\Delta{\cal E}(a) -\frac{\pi^2}{1440 (a/2)^3},
\end{equation}
which immediately leads to the Casimir interaction energy per unit area
for the complete stack in FIG.~\ref{fig-a2i-plates-12} given by
\begin{equation}
\Delta{\cal E}(a) = +\frac{8}{7}\frac{\pi^2}{1440 a^3}.
\label{cie-ingp}
\end{equation}
Thus, using the idea of self-similarity, in a self-contained derivation,
we have derived the Casimir interaction energy of an infinite stack 
of plates. Remarkably, the sign of the Casimir interaction energy for 
this configuration is positive. Thus, the tendency
for the infinite sequence of plates in FIG.~\ref{fig-a2i-plates-12}
is to inflate due to the pressure of vacuum.

\begin{figure*}
\begin{center}
\includegraphics{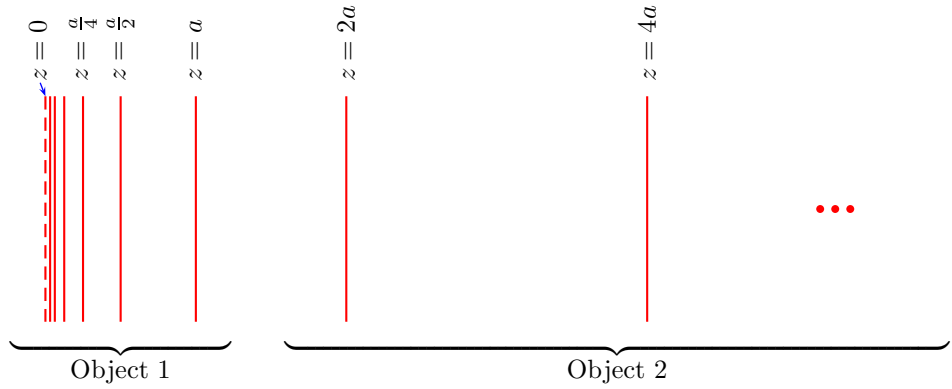}
\caption{A geometric sequence of parallel $\delta$-function plates.
The position of the plates is given by the sequence
$z= \ldots, \frac{a}{8}, \frac{a}{4}, \frac{a}{2}, a, 2a,4a,8a,\dots$.
}
\label{fig-a2i-plates-double}
\end{center}
\end{figure*}%

We consider another example
to point out that the Casimir interaction energy
is not always positive for an infinite sequence of plates.
We consider an infinite sequence of plates placed at the following positions:
\begin{equation}
z=2a,4a,8a,16a,\dots,
\label{ser-gp-in}
\end{equation}
as described in FIG.~\ref{fig-a2i-plates-in}.
(We start from $z=2a$ because it extends the series in Eq.~(\ref{zpos-gpd})
and later allows us to merge the two stacks.)
Using the idea of self-similarity, we identify the relation
\begin{equation}
\Delta{\cal E}(2a) = \Delta{\cal E}(4a) -\frac{\pi^2}{1440 (2a)^3}.
\end{equation}
Then, using
\begin{equation}
\Delta{\cal E}(4a) = \frac{1}{2^3}\,\Delta{\cal E}(2a), 
\end{equation}
we immediately learn that
\begin{equation}
\Delta{\cal E}(2a) = -\frac{1}{7}\frac{\pi^2}{1440 a^3}.
\label{cie-degp}
\end{equation}
This suggests that the tendency of the stack of plates 
in FIG.~\ref{fig-a2i-plates-in} is to contract under the pressure of vacuum.

Having derived the Casimir interaction energy of two independent stacks,
we now place them such that they can be imagined to be
a sequence that extends on both ends, given by
\begin{equation}
z= \ldots, \frac{a}{8}, \frac{a}{4}, \frac{a}{2}, a, 2a,4a,8a,\dots,
\label{zpos-gpd-do}
\end{equation}
as described in FIG.~\ref{fig-a2i-plates-double}.
Since we already derived the energies for the individual stacks,
we can calculate the energy of the complete stack using the two-body
break up of the Casimir energies. Thus, we have the total interaction energy
of the two stacks $\Delta{\cal E}_\text{tot}(a)$ given by the relation
\begin{equation}
\Delta{\cal E}_\text{tot}(a) = \Delta{\cal E}(a) + \Delta{\cal E}(2a)
-\frac{\pi^2}{1440 a^3},
\end{equation}
where the first term on the right is the Casimir interaction energy
$\Delta{\cal E}(a)$ of the first stack given by Eq.~(\ref{cie-ingp}),
the second term  is the Casimir interaction energy
$\Delta{\cal E}(2a)$ of the second stack given by Eq.~(\ref{cie-degp}),
and the third term is the interaction energy of the two stacks
given by the energy of two Dirichlet plates in Eq.~(\ref{Ce-twop}).
Together we have
\begin{equation}
\Delta{\cal E}_\text{tot}(a) = +\frac{8}{7} \frac{\pi^2}{1440 a^3}
-\frac{1}{7} \frac{\pi^2}{1440 a^3} -\frac{\pi^2}{1440 a^3} =0,
\end{equation}
which suggests that the Casimir energy of the two stacks, in
conjunction, in FIG.~\ref{fig-a2i-plates-double},
is exactly zero. Apparently, the pressure due to vacuum that tends
to inflate the first stack, when in isolation, 
and contract the second stack in isolation, when in conjunction,
conspire to balance these opposite tendencies exactly.
It can be easily verified that this cancellation is independent
of the particular choice of breakup into Objects 1 and 2,
which is a signature of self-similarity.

\begin{figure}[ht]
\begin{center}
\includegraphics{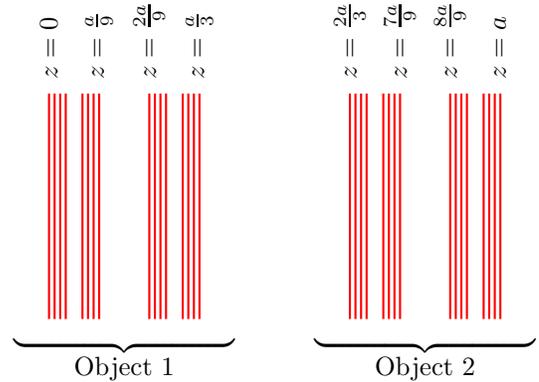}
\caption{A sequence of parallel plates positioned at the points
forming the Cantor set. The figure shows plates positioned at points
generated in four iterations.
}
\label{fig-cantor-plates}
\end{center}
\end{figure}%
In our last example, we highlight  
a self-similar configuration of plates motivated from 
the Cantor set. We place a $\delta$-function plate at every point
of the Cantor set. The classic Cantor set is obtained by 
iteratively dividing a line segment into three parts and deleting
the central region each time.
We build our stack of plates by placing a $\delta$-function plate
at the edge of the remaining segments in each iteration,
see FIG.~\ref{fig-cantor-plates}.
The idea of self-similarity and the two-body break up of energy then 
leads to the relation, in the Dirichlet limit,
\begin{equation}
\Delta{\cal E}(a) = \Delta{\cal E}(a/3) + \Delta{\cal E}(a/3) 
-\frac{\pi^2}{1440 (a/3)^3}.
\end{equation}
Then, using
\begin{equation}
\Delta{\cal E}(a/3) = 3^3\,\Delta{\cal E}(a)
\end{equation}
we have the Casimir interaction for the configuration in 
FIG.~\ref{fig-cantor-plates} given by
\begin{equation}
\Delta{\cal E}(a) = +\frac{27}{53}\frac{\pi^2}{1440 a^3}.
\end{equation}
The positive sign signifies that the pressure due to vacuum tends to 
inflate the stack in FIG.~\ref{fig-cantor-plates}.

In the following section, we further support the above derivations for the 
Casimir interaction energies for self-similar plates by evaluating 
the explicit Green's functions for these configurations.
We find the Green's function for
$N$ parallel $\delta$-function plates given in terms of the corresponding
combinations of $(N-2)$ parallel $\delta$-function plates.
We are interested in the limit of infinite plates
obtained by taking the limit $N\to\infty$.

\section{Green's function for $N$ parallel $\delta$-function plates}
\label{sec-Gfuns}

The Green's function for $N$ parallel $\delta$-function plates
satisfies the equation
\begin{equation}
\left[ -\frac{d^2}{dz^2} +\kappa^2 +\sum_{i=1}^N\lambda_i \delta(z-a_i) 
\right] g_{1\ldots N}(z,z^\prime) = \delta(z-z^\prime).
\label{gf-Ndplates}
\end{equation}
The translation symmetry in the plane of plates and static considerations
allows the corresponding modes to be bunched as $\kappa^2=k_\perp^2-\omega^2$.
We also make Euclidean rotation and replace $\omega=i\zeta$.
The free Green's function, corresponding to the absence of all the plates,
satisfies the equation
\begin{equation}
\left( -\frac{d^2}{dz^2} +\kappa^2 \right)
g_0(z-z^\prime) = \delta(z-z^\prime),
\label{gf-free}
\end{equation}
and has the solution
\begin{equation}
g_0(z-z^\prime) = \frac{1}{2\kappa} e^{-\kappa |z-z^\prime|}.
\end{equation}
We use the ansatz
\begin{equation}
g_{1\ldots N}(z,z^\prime) = g_0(z,z^\prime) 
- {\bf r}(z) \cdot {\bf t}_{1\ldots N} \cdot {\bf r}(z^\prime),
\label{ans-Ngf}
\end{equation}
which is motivated from the discussions in Ref.\,\cite{Shajesh:2011ef}.
In Eq.~(\ref{ans-Ngf}), we have used matrix notation and summation
convention to symbolically write
\begin{equation}
{\bf r}(z) \cdot {\bf t}_{1\ldots N} \cdot {\bf r}(z^\prime)
=r_i(z) \, t^{ij}_{1\ldots N} \, r_j(z^\prime),
\end{equation}
where the components $r_i(z)$ of the vector ${\bf r}(z)$ are free
Green's functions when one of the source points is on the $i$-th plate, that is,
\begin{equation}
r_i(z) = g_0(z-a_i) = \frac{1}{2\kappa} e^{-\kappa |z-a_i|}.
\label{ri-def}
\end{equation}
The components $t_{1\ldots N}^{ij}$ of the dyadic ${\bf t}_{1\ldots N}$ 
are independent of $z$ and $z^\prime$ and are given by the matrix equation,
see the Appendix\,\ref{sec-Faddeev-Eqn-proof},
\begin{equation}
{\bf t}_{1\ldots N} = \left( {\bf 1} + {\bm\lambda} \cdot {\bf R} \right)^{-1} 
\cdot {\bm\lambda},
\label{tm-sol}
\end{equation}
where ${\bf 1}$ is the identity matrix,
\begin{equation}
{\bm\lambda} = \left[ \begin{array}{cccc}
\lambda_1 &&& \rput(-0.3,-0.3){\text{\huge 0}} \\
&\lambda_2  && \\ && \ddots & \\
\rput(0.3,0.3){\text{\huge 0}} &&& \lambda_N
\end{array} \right]
\end{equation}
is a diagonal matrix of coupling constants, and
\begin{equation}
{\bf R} = \left[ \begin{array}{cccc}
g_0(0) & g_0(a_1-a_2) & \ldots & g_0(a_1-a_N) \\
g_0(a_2-a_1) &g_0(0) & \ldots & g_0(a_2-a_N) \\
\vdots & \vdots & \ddots & \vdots \\
g_0(a_N-a_1) &g_0(a_N-a_2) & \ldots & g_0(0)
\end{array} \right]
\label{Rm-def}
\end{equation}
is a matrix whose components $R_{ij}$ are free Green's functions
evaluated from the $i$-th to the $j$-th plate. That is,
\begin{equation}
R_{ij} =g_0(a_i-a_j)= \frac{1}{2\kappa} e^{-\kappa |a_i-a_j|}.
\label{Rij-not}
\end{equation}
For convenience, we introduce dimensionless quantities
\begin{equation}
\tilde {\bm\lambda} = \frac{{\bm\lambda}}{2\kappa}, \quad
\tilde {\bf t}_{1\ldots N} = \frac{{\bf t}_{1\ldots N}}{2\kappa},
\quad \text{and} \quad \tilde {\bf R} = 2\kappa {\bf R}.
\label{dimless-def}
\end{equation}
The matrix equations of Eq.~(\ref{tm-sol}) are the
Faddeev equations~\cite{Faddeev:1965a, Merkuriev:1993a}
that were introduced in the study of nuclear many-body scattering.

For a single plate, $N=1$, we immediately have
\begin{equation}
\tilde t_1 = \frac{\tilde\lambda_1}{1+\tilde\lambda_1}.
\end{equation}
The corresponding Green's function, given by Eq.~(\ref{ans-Ngf}) for $N=1$,
has the explicit form
\begin{equation}
g_1(z,z^\prime) = \frac{1}{2\kappa}e^{-\kappa |z-z^\prime|}
- \frac{\tilde t_1}{2\kappa} e^{-\kappa |z-a_1|} e^{-\kappa |z^\prime-a_1|}.
\label{gi-sb-def}
\end{equation}
The solution for the Green's function in Eq.~(\ref{gi-sb-def}) is 
valid for all $z$ and $z^\prime$, the difference in the behavior 
decided by the absolute values $|z-a_1|$ and $|z^\prime-a_1|$.
The compactness in the solution for planar geometry is a direct 
consequence of this feature, which does not extend to other geometries.

For two plates, $N=2$, we solve Eq.~(\ref{tm-sol}) and find 
\begin{equation}
\tilde {\bf t}_{12} = \frac{1}{\Delta_{12}}
\left[ \begin{array}{cc}
\tilde t_1 & -\tilde t_1 \tilde R_{12} \tilde t_2 \\
-\tilde t_2 \tilde R_{21} \tilde t_1 & \tilde t_2 
\end{array} \right],
\end{equation}
where
\begin{equation}
\Delta_{12} = 1 - \tilde t_1 \tilde R_{12} \tilde t_2 \tilde R_{21}.
\end{equation}
\begin{widetext}
The corresponding Green's function, given by Eq.~(\ref{ans-Ngf}) for $N=2$,
has the explicit form
\begin{eqnarray}
g_{12}(z,z^\prime) &=& \frac{1}{2\kappa} e^{-\kappa|z-z^\prime|}
- \frac{1}{2\kappa} \frac{1}{\Delta_{12}} \text{Tr} \left[ \begin{array}{cc}
\tilde t_1 & -\tilde t_1 \tilde R_{12} \tilde t_2 \\
-\tilde t_2 \tilde R_{21} \tilde t_1 & \tilde t_2
\end{array} \right] 
\left[ \begin{array}{cc}
e^{-\kappa|z-a_1|} e^{-\kappa|z^\prime-a_1|} 
& e^{-\kappa|z-a_1|} e^{-\kappa|z^\prime-a_2|} \\
e^{-\kappa|z-a_2|} e^{-\kappa|z^\prime-a_1|} & 
e^{-\kappa|z-a_2|} e^{-\kappa|z^\prime-a_2|} \end{array} \right],
\label{gij-tb-def}
\end{eqnarray}
where we used the property of trace to write the second term in 
Eq.~(\ref{ans-Ngf}) in the form
\begin{equation}
{\bf r}(z) \cdot {\bf t}_{1\dots N} \cdot {\bf r}(z^\prime)
= \text{Tr} \Big[ \,{\bf t}_{1\dots N} 
\cdot {\bf r}(z^\prime)\, {\bf r}(z)^T \Big].
\end{equation}

For three plates, $N=3$, we solve Eq.~(\ref{tm-sol}) and find 
\begin{equation}
\tilde {\bf t}_{123} = \frac{1}{\Delta_{123}}
\left[ \begin{array}{ccc}
\tilde t_1 \Delta_{23} & -\tilde t_1 \tilde R_{1[3]2} \tilde t_2 
& -\tilde t_1 \tilde R_{1[2]3} \tilde t_3 \\
-\tilde t_2 \tilde R_{2[3]1} \tilde t_1 & \tilde t_2 \Delta_{13} 
& -\tilde t_2 \tilde R_{2[1]3} \tilde t_3 \\
-\tilde t_3 \tilde R_{3[2]1} \tilde t_1 &
-\tilde t_3 \tilde R_{3[1]2} \tilde t_2 & \tilde t_3 \Delta_{12} 
\end{array} \right],
\end{equation}
where the determinant $\Delta_{123}$ can be written in the form
\begin{equation}
\Delta_{123} = \Delta_{23} 
- \tilde t_1 \tilde R_{12} \tilde t_2 \tilde R_{2[3]1}
- \tilde t_1 \tilde R_{13} \tilde t_3 \tilde R_{3[2]1}.
\end{equation}
Here, we have introduced the generalized form of the notation in 
Eq.~(\ref{Rij-not}),
\begin{equation}
R_{i[k]j} = g_k(a_i,a_j),
\end{equation}
the right side of which are given in terms of 1-plate Green's functions
of Eq.~(\ref{gi-sb-def}).
The corresponding Green's function $g_{123}(z,z^\prime)$
is given by Eq.~(\ref{ans-Ngf}) for $N=3$,
the second term of which has the explicit form
\begin{eqnarray}
- \frac{1}{2\kappa} \frac{1}{\Delta_{123}} \text{Tr} \left[ \begin{array}{ccc}
\tilde t_1 \Delta_{23} & -\tilde t_1 \tilde R_{1[3]2} \tilde t_2
& -\tilde t_1 \tilde R_{1[2]3} \tilde t_3 \\
-\tilde t_2 \tilde R_{2[3]1} \tilde t_1 & \tilde t_2 \Delta_{13}
& -\tilde t_2 \tilde R_{2[1]3} \tilde t_3 \\
-\tilde t_3 \tilde R_{3[2]1} \tilde t_1 &
-\tilde t_3 \tilde R_{3[1]2} \tilde t_2 & \tilde t_3 \Delta_{12}
\end{array} \right] 
\left[ \begin{array}{ccc}
e^{-\kappa|z-a_1|} e^{-\kappa|z^\prime-a_1|}
& e^{-\kappa|z-a_1|} e^{-\kappa|z^\prime-a_2|}
& e^{-\kappa|z-a_1|} e^{-\kappa|z^\prime-a_3|} \\
e^{-\kappa|z-a_2|} e^{-\kappa|z^\prime-a_1|}
& e^{-\kappa|z-a_2|} e^{-\kappa|z^\prime-a_2|} 
& e^{-\kappa|z-a_2|} e^{-\kappa|z^\prime-a_3|} \\
e^{-\kappa|z-a_3|} e^{-\kappa|z^\prime-a_1|}
& e^{-\kappa|z-a_3|} e^{-\kappa|z^\prime-a_2|} 
& e^{-\kappa|z-a_3|} e^{-\kappa|z^\prime-a_3|} \end{array} \right].
\hspace{8mm}
\end{eqnarray}

For $N=4$, we solve Eq.~(\ref{tm-sol}) and find
\begin{equation}
\tilde {\bf t}_{1234} = \frac{1}{\Delta_{1234}}
\left[ \begin{array}{cccc}
\tilde t_1 \Delta_{234} & -\tilde t_1 \tilde R_{1[34]2} \tilde t_2 \Delta_{34}
& -\tilde t_1 \tilde R_{1[24]3} \tilde t_3 \Delta_{24}
& -\tilde t_1 \tilde R_{1[23]4} \tilde t_4 \Delta_{23} \\
-\tilde t_2 \tilde R_{2[34]1} \tilde t_1 \Delta_{34} & \tilde t_2 \Delta_{134}
& -\tilde t_2 \tilde R_{2[14]3} \tilde t_3 \Delta_{14}
& -\tilde t_2 \tilde R_{2[13]4} \tilde t_4 \Delta_{13} \\
-\tilde t_3 \tilde R_{3[24]1} \tilde t_1 \Delta_{24} &
-\tilde t_3 \tilde R_{3[14]2} \tilde t_2 \Delta_{14} & \tilde t_3 \Delta_{124}
& -\tilde t_3 \tilde R_{3[12]4} \tilde t_4 \Delta_{12} \\
-\tilde t_4 \tilde R_{4[23]1} \tilde t_1 \Delta_{23} &
-\tilde t_4 \tilde R_{4[13]2} \tilde t_2 \Delta_{13} &
-\tilde t_4 \tilde R_{4[12]3} \tilde t_3 \Delta_{12} & \tilde t_4 \Delta_{123}
\end{array} \right],
\end{equation}
where
\begin{equation}
R_{i[mn]j} = g_{mn}(a_i,a_j),
\end{equation}
the right side of which are given in terms of 2-plate Green's functions
of Eq.~(\ref{gij-tb-def}). The determinant 
\begin{equation}
\Delta_{1234} = \Delta_{234}
- \tilde t_1 \tilde R_{12} \tilde t_2 \tilde R_{2[34]1} \Delta_{34}
- \tilde t_1 \tilde R_{13} \tilde t_3 \tilde R_{3[24]1} \Delta_{24}
- \tilde t_1 \tilde R_{14} \tilde t_4 \tilde R_{4[23]1} \Delta_{23}.
\end{equation}

\subsection{Recursion relation}

From the pattern that emerges for the above $N=1,2,3,4$ cases,
we can write down the the Green's function for $N$ $\delta$-function plates as
\begin{equation}
\tilde {\bf t}_{12\ldots N} = \frac{1}{\Delta_{12\ldots N}}
\left[ \begin{array}{cccc}
\tilde t_1 \Delta_{23\ldots N} 
& -\tilde t_1 \tilde R_{1[34\ldots N]2} \tilde t_2 \Delta_{34\ldots N}
& \cdots 
& -\tilde t_1 \tilde R_{1[23\ldots N-1]N} \tilde t_N \Delta_{23\ldots N-1} 
\\[2mm]
-\tilde t_2 \tilde R_{2[34\ldots N]1} \tilde t_1 \Delta_{34\ldots N} 
& \tilde t_2 \Delta_{134\ldots N} & \cdots 
& -\tilde t_2 \tilde R_{2[13\ldots N-1]N} \tilde t_N \Delta_{13\ldots N-1} 
\\[2mm] \vdots & \vdots & \ddots & \vdots \\[2mm]
-\tilde t_N \tilde R_{N[23\ldots N-1]1} \tilde t_1 \Delta_{23\ldots N-1} &
-\tilde t_N \tilde R_{N[13\ldots N-1]2} \tilde t_2 \Delta_{13\ldots N-1} &
\cdots & \tilde t_N \Delta_{12\ldots N-1}
\end{array} \right],
\end{equation}
This can then be immediately extended for the $N\to\infty$ case. 
The Green's function for $N$ plates is given in terms of 
all possible Green's function for $(N-2)$ plates, obtained
by deleting two plates. In this sense, we have a recursion relation
for the Green's function. 
The Green's function presented as a recursion relation is very suitable
for the kind of problems we are addressing here.
Our method for finding the Green's function for $N$ bodies is fundamentally
different from the earlier techniques used to find the 
Green's function for multilayered systems,
for example, see Refs.~\cite{Reed1987gm,Tomas1995cp,Zhou1995wm}.

\subsection{Green's function for a sequence of Dirichlet plates}

Let us consider the very special case of every $\delta$-function plate
being a Dirichlet plate. This is described by the limiting conditions,
\begin{equation}
\lambda_i \to \infty,
\end{equation}
for all $i$'s. We go back to the matrix equation in Eq.~(\ref{tm-sol})
and find that in the Dirichlet limit (in all plates) we have 
\begin{equation}
{\bf t}_{1\ldots N} = {\bf R}^{-1},
\end{equation}
where ${\bf R}$ was defined in Eq.~(\ref{Rm-def}) and is a matrix
built out of all possible free Green's functions. 
The inverse of ${\bf R}$ is immediately evaluated to yield
the transition matrix as a tridiagonal matrix,
\begin{equation}
{\bf t}_{1\ldots N} = \kappa
\left[ \begin{array}{ccccccc}
D_{11} & S_{12} &&&&& \rput(-1.5,-0.9){\text{\Huge 0}} \\[2mm]
S_{21} & D_{22} & S_{23} &&&& \\[2mm]
& S_{32} & D_{33} & S_{34} &&& \\[0.5mm]
&& S_{43} & D_{44} & \ddots && \\[1mm]
&&&\ddots & \ddots & \ddots \hspace{4mm} & \\[0mm]
&&&&\ddots & D_{N-1,N-1} & S_{N-1,N} \\[2mm]
\rput(1.0,0.9){\text{\Huge 0}} &&&&& S_{N,N-1} & D_{N,N}
\end{array} \right],
\label{tm-Dp-sol}
\end{equation}
where
\begin{equation}
D_{11} = \frac{e^{\kappa a_{12}}}{\sinh\kappa a_{12}},
\qquad D_{NN} = \frac{e^{\kappa a_{N-1,N}}}{\sinh\kappa a_{N-1,N}},
\end{equation}
and
\begin{equation}
D_{ii} = \frac{e^{\kappa a_{i-1,i}}}{\sinh\kappa a_{i-1,i}}
-2 + \frac{e^{\kappa a_{i,i+1}}}{\sinh\kappa a_{i,i+1}},
\qquad \text{if} \qquad i\neq 1, i\neq N,
\end{equation}
and
\begin{equation}
S_{i,i+1}=S_{i+1,i} =
-\frac{1}{\sinh\kappa a_{i,i+1}},
\end{equation}
such that $a_{ij}$ is the magnitude of the distance between 
the $i$-th and $j$-th parallel plate.
The Green's function is then completely determined by the transition matrix
using Eq.~(\ref{ans-Ngf}).

\section{Casimir energy for $N$ parallel $\delta$-function plates}
\label{sec-CaseI}. 

The Casimir energy per unit area for $N$ parallel $\delta$-function plates
is determined in terms of the Green's function \cite{Shajesh:2011ef},
\begin{equation}
{\cal E}_{1\ldots N} = -\frac{1}{6\pi^2} \int_0^\infty \kappa^4d\kappa
\int_{-\infty}^\infty dz\,g_{1\dots N}(z,z).
\label{NpCas}
\end{equation}
Using the ansatz of Eq.~(\ref{ans-Ngf}) in Eq.~(\ref{NpCas}), we have
\begin{equation}
{\cal E}_{1\ldots N}
= {\cal E}_0 +\frac{1}{6\pi^2} \int_0^\infty \kappa^4d\kappa
\,\text{Tr} \,{\bf t}_{1\ldots N} \cdot
\int_{-\infty}^\infty dz\,{\bf r}(z) {\bf r}(z)^T,
\label{en-e0Cep}
\end{equation}
where ${\cal E}_0$ is the energy in the absence of all plates,
a divergent quantity, often called the bulk free energy, given by
\begin{equation}
{\cal E}_0 = -\frac{1}{6\pi^2} \int_0^\infty \kappa^4d\kappa
\int_{-\infty}^\infty dz\,g_0(z,z).
\end{equation}
Observing that $g_0(z,z)=1/2\kappa$, one notes that this 
divergent contribution is proportional to the volume 
$A\int_{-\infty}^\infty dz$ and independent of any of the 
physical parameters of the problem. For planar geometries that we are 
discussing the $z$-integral in Eq.~(\ref{en-e0Cep}) can be evaluated to yield
\begin{equation}
\int_{-\infty}^\infty dz\, {\bf r}(z) {\bf r}(z)^T
= \frac{{\bf M}}{4\kappa^3},
\end{equation}
where
\begin{equation}
{\bf M} =
\left[ \begin{array}{cccc}
1 & (1+\kappa a_{12})\,e^{-\kappa a_{12}} & \cdots
& (1+\kappa a_{1N})\,e^{-\kappa a_{1N}} \\[1mm]
(1+\kappa a_{21})\,e^{-\kappa a_{21}} &1 & \cdots
& (1+\kappa a_{2N})\,e^{-\kappa a_{2N}} \\[1mm]
\vdots & \vdots & \ddots & \vdots \\[1mm]
(1+\kappa a_{N1})\,e^{-\kappa a_{N1}} & (1+\kappa a_{N2})\,e^{-\kappa a_{N2}} 
& \cdots & 1 
\end{array} \right],
\label{pp-nmat}
\end{equation}
where $a_{ij}$ is the distance between the $i$-th and $j$-th parallel plate,
defined previously. Thus, we have the Casimir energy per unit area given by
\begin{equation}
{\cal E}_{1\dots N} = {\cal E}_0
+ \frac{1}{24\pi^2} \int_0^\infty \kappa d\kappa
\,\text{Tr} \,{\bf t}_{1\ldots N} \cdot {\bf M}.
\label{cen-ine28}
\end{equation}
Other than the bulk free energy term, ${\cal E}_0$, we also have divergent 
single-body contributions from each of the $N$ individual plates. 
The Casimir energy of a single plate, say plate 1,
in the absence of all other plates, is given using Eq.~(\ref{cen-ine28}) as
\begin{equation}
{\cal E}_1 = {\cal E}_0 +\frac{1}{12\pi^2} \int_0^\infty \kappa^2d\kappa\,
\tilde t_1,
\label{sinB-ce}
\end{equation}
which is divergent and independent of any of the physical parameters
of the problem. The second term in Eq.~(\ref{cen-ine28}) has $N$
divergent contributions of these, and we study the contribution
to the energy after these single-body contributions, in addition to the
bulk free energy, has been subtracted. To this end we define
the Casimir interaction energy per unit area, because they involve
interactions between the plates, in the spirit of Eq.~(\ref{en-mbbu}),
\begin{equation}
\Delta {\cal E}_{1\ldots N} = {\cal E}_{1\ldots N}
- {\cal E}_0 - \sum_{i=1}^N \Delta{\cal E}_i,
\label{cas-in-e}
\end{equation}
which is free of divergences unless any of the individual plates touch.

\subsection{Finite sequence of Dirichlet plates}

For a sequence of Dirichlet plates the Casimir energy is given by
\begin{equation}
{\cal E}_{1\ldots N} 
= {\cal E}_0 + \frac{1}{12\pi^2} \int_0^\infty \kappa^2d\kappa
\left[ -(N-2) + \sum_{i=1}^{N-1} \frac{ \left[ e^{\kappa a_{i,i+1}} 
-(1+\kappa a_{i,i+1}) e^{-\kappa a_{i,i+1}} \right]}
 {\sinh \kappa a_{i,i+1}} \right], 
\label{cas-Dpl-12}
\end{equation}
where the contribution of $(N-2)$ inside the square brackets comes
from summing the $-2$'s in the diagonal terms of the transition matrix in
Eq.~(\ref{tm-Dp-sol}). This term and the first two terms inside the 
sum in Eq.~(\ref{cas-Dpl-12}) can be combined as
\begin{equation}
\frac{1}{12\pi^2} \int_0^\infty \kappa^2d\kappa \Big[-(N-2) + 2(N-1) \Big]
= \frac{N}{12\pi^2} \int_0^\infty \kappa^2d\kappa, 
\end{equation}
which is identified as the sum of single-body (divergent) contributions
to the Casimir energy from the $N$ individual plates, see Eq.~(\ref{sinB-ce}).
Thus, the Casimir interaction energy per unit area, 
introduced in Eq.~(\ref{cas-in-e}),
for $N$ parallel Dirichlet plates is given by the expression
\begin{equation}
\Delta {\cal E}_{1\ldots N}
=-\frac{1}{12\pi^2} \sum_{i=1}^{N-1} \int_0^\infty \kappa^2 d\kappa
\frac{\kappa a_{i,i+1} e^{-\kappa a_{i,i+1}}}{\sinh\kappa a_{i,i+1}},
\end{equation}
which using the integral
\begin{equation}
\int_0^\infty \frac{x^3dx\,e^{-x}}{\sinh x} = \frac{\pi^4}{1440}, 
\end{equation} 
is expressed as
\begin{equation}
\Delta {\cal E}_{1\ldots N} 
= -\frac{\pi^2}{1440} \sum_{i=1}^{N-1} \frac{1}{a_{i,i+1}^3}.
\label{ceD-genE}
\end{equation}
This result is not surprising because a Dirichlet plate physically
disconnects the two half-spaces across it.
\end{widetext}

\subsection{Infinite sequence of Dirichlet plates}

In the example of FIG.~\ref{fig-a2i-plates-12} given by
the sequence of plates in Eq.~(\ref{zpos-gpd})
we have
\begin{equation}
a_{i,i+1} = \frac{a}{2^i},
\end{equation}
which together with Eq.~(\ref{ceD-genE}) leads to
the Casimir interaction energy for this configuration
given by the expression
\begin{equation}
\Delta{\cal E}_{12\ldots} = -\frac{\pi^2}{1440 a^3} (8+8^2+8^3+\dots).
\end{equation}
Using the idea of self-similarity in the context of series we identify the
relation $x=8+8x$, with $x=8+8^2+8^3+\dots$.
Thus, we make the formal assignment
\begin{equation}
8+8^2+8^3+\dots = -\frac{8}{7}
\end{equation}
to determine the the Casimir interaction energy for this configuration
to be
\begin{equation}
\Delta{\cal E}_{12\ldots} = \frac{8}{7} \frac{\pi^2}{1440 a^3},
\end{equation}
exactly as we derived earlier in Eq.~(\ref{cie-ingp}).

Next, we consider an infinite sequence of Dirichlet plates
given using Eq.(\ref{ser-gp-in}), as described in 
FIG.~\ref{fig-a2i-plates-in}, such that
\begin{equation}
a_{i,i+1} = 2^ia.
\end{equation}
We have the Casimir interaction energy for this configuration,
using Eq.~(\ref{ceD-genE}), given by the expression
\begin{equation}
\Delta{\cal E}_{12\ldots} = -\frac{\pi^2}{1440 a^3}
\left(\frac{1}{8}+ \frac{1}{8^2}+ \frac{1}{8^3} +\dots \right).
\end{equation}
This involves the convergent series
\begin{equation}
\frac{1}{8}+ \frac{1}{8^2}+\frac{1}{8^3}+\dots = \frac{1}{7},
\end{equation}
which implies that
the Casimir interaction energy for this configuration is
\begin{equation}
\Delta{\cal E}_{12\ldots} = -\frac{1}{7} \frac{\pi^2}{1440 a^3}
\end{equation}
exactly as we derived earlier in Eq.~(\ref{cie-degp}).

\begin{figure}[ht]
\begin{center}
\includegraphics{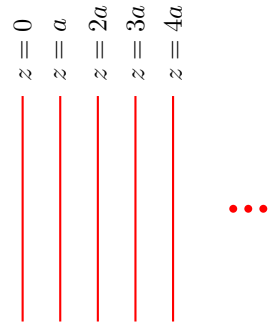}
\caption{Equidistant identical $\delta$-function plates 
filling half of the space. }
\label{fig-equidistant-plates-hs}
\end{center}
\end{figure}%
As the final example, we consider equidistant
Dirichlet plates filling half of the space, that is,
\begin{equation}
z=0,a,2a,3a,\ldots,
\end{equation}
such that
\begin{equation}
a_{i,i+1} = a,
\end{equation}
for all $i$, see FIG.~\ref{fig-equidistant-plates-hs}.
We have the Casimir interaction energy for this configuration
given by the expression
\begin{equation}
\Delta{\cal E}_{12\ldots} = -\frac{\pi^2}{1440 a^3} (1+1+1+ \dots).
\end{equation}
If we now make the formal assignment 
\begin{equation}
1+1+1+\dots = \zeta(0) = -\frac{1}{2},
\end{equation}
the Casimir interaction energy of this infinite equidistant Dirichlet plates 
filling half of the space changes sign,
\begin{equation}
\Delta{\cal E}_{12\ldots} = \frac{1}{2} \frac{\pi^2}{1440 a^3}.
\end{equation}
That is, again, the tendency for the plates is to inflate under the 
pressure of vacuum. 

\section{Analogy to the theory of the piecewise uniform string}
\label{sec-rel-string}

There exists an interesting analogy between the theory considered 
in this paper and Casimir theory of the piecewise uniform string.
To our knowledge, this analogy has not been pointed out before. Let us 
start by outlining some basic aspects of this kind of string theory, 
assuming first that the system is a material ring of total length $L$ 
divided into two pieces, $L=L_1+L_2$. The system exhibits small 
oscillations with amplitude $\psi(\sigma, \tau)$, where $\sigma$ is  
the position coordinate and $\tau$ the time (the usual convention in 
string theory). The string tensions are $T_1$ and $T_2$, and the 
mass densities are $\rho_1$ and $\rho_2$, adjusted such that the 
speed of sound $\sqrt{T/\rho}$ is everywhere the same as the 
speed of light,
\begin{equation}
\sqrt{\frac{T_1}{\rho_1}}=\sqrt{\frac{T_2}{\rho_2}}=1.
\label{X0}
\end{equation}
In this sense, the string model is relativistic. At the two junctions, 
the displacement $\psi$, as well as the transverse force 
$T\partial \psi/\partial \sigma$, are continuous. The equation of motion
\begin{equation}
\left( \frac{\partial^2}{\partial \sigma^2}
-\frac{\partial^2}{\partial \tau^2}\right) \psi=0
\end{equation}
is solved  for the right-and left-moving modes. 
The dispersion relation determining the eigenfrequencies is
\begin{equation}
\frac{4x}{(1-x)^2}\sin^2\left( \frac{\omega L}{2}\right) 
+\sin \omega L_1 \sin \omega L_2=0, \label{X1}
\end{equation}
where $x=T_1/T_2$ is the tension ratio.

The Casimir energy, given by the difference between the total energy 
$E_{1+2}$ and the energy $E_\text{uniform}$ for a uniform string, is
\begin{equation}
E=E_{1+2}-E_\text{uniform}=\frac{1}{2}\sum \omega_n-E_\text{uniform}.
\end{equation}
It can be regularized in at least three different ways:
\begin{itemize}
\item Use of a cutoff factor $f=e^{-\alpha \omega_n}$, $\alpha \ll 1$, 
being applied to the energy expression before summing over the modes.
\item Use of the contour integration method, which means applying 
the so-called argument principle
\begin{equation}
\frac{1}{2\pi i} \oint \omega \frac{d}{d\omega}\ln g(\omega)d\omega 
=\sum \omega_0-\sum \omega_\infty,
\end{equation}
which holds for any meromorphic function $g(\omega)$, where $\omega_0$
and $\omega_\infty$ denote the zeros and the poles, respectively. 
In our case, $g(\omega)$ is essentially the left-hand side of the 
expression in Eq.~(\ref{X1}) above.
\item Use of the zeta-function method, which in our case means 
applying the analytic continuation of the Hurwitz function 
$\zeta_H(s,a)$ defined as
\begin{equation}
\zeta_H(s,a)=\sum_{n=0}^\infty (n+a)^{-s}, \quad 0<a<1, \quad \Re\, s>1.
\end{equation}
\end{itemize}
All methods lead to the same answer for the Casimir energy, 
due to the relativistic property in Eq.~(\ref{X0}). We give the expression 
only for the simple case when $x \rightarrow 0$,
\begin{equation}
E=-\frac{\pi}{24L}\left( \frac{L_2}{L_1}+\frac{L_1}{L_2}-2\right).
\end{equation}
The energy is seen to be zero (if $L_1=L_2$) or otherwise negative.
The difference in the coefficient relative to Eq.~(\ref{Ce-twop})
is because we are working in 1+1 spacetime dimensions here as compared
to 3+1 spacetime dimensions earlier.

To our knowledge, this model was first suggested by Brevik and Nielsen 
in \cite{brevik90}, c.f. also Li et al. \cite{li91}, 
applying the Hurwitz zeta function. The contour integration method 
was applied to this problem by Brevik and Elizalde \cite{brevik94}. 
Later on, there have been developments in various directions, 
including the generalization to a string composed of $2N$ pieces, 
all of the same length \cite{brevik95}. General reviews, 
containing more references, can be found 
in Refs.~\cite{brevik03,berntsen97,elizalde95}.
A generalization to the case of a nonrelativistic string, 
(the velocity of sound being different in the different pieces,)
has been given in Ref.~\cite{hadasz00}.

We are now in a position to see the natural relationship to the 
model with self-similar plates. Consider the case where the positions 
are given by $z=a, a/2, a/4, a/8,\dots$. The difference between the 
positions of the first and last plate in the limit where the number 
of plates is infinity is finite, equal to $a$. Assume now that the 
composite string is divided into alternating type 1 and type 2 sections,
spaced according to the same prescription. This means simply that the 
string length $L$ is to be identified with $a$. It would be  of interest 
to carry out a calculation of the Casimir energy for this special 
kind of string. We do not enter into this task here, however, 
but limit ourselves to pointing out the analogy.

The relativistic property of the system will still be maintained, 
due to  Eq.~(\ref{X0}), although an evident physical restriction 
is that the dielectric property cannot be maintained of the elements 
when their lengths go towards zero. The limit of infinitely many 
pieces is an idealized model.

\section{Conclusions and outlook}
\label{sec-con-out}

We have derived the Casimir energies of simple self-similar
configurations consisting of parallel $\delta$-function plates
satisfying Dirichlet boundary conditions
using the idea of self-similarity alone.
Then, we have corroborated our results for Casimir energies
using the completely independent Green's functions formalism.
We have thus shown that an infinite stack of parallel plates 
can have positive, negative, or zero Casimir energy.
In particular, we have successfully derived the Casimir energy of a stack of 
plates positioned at the points of the Cantor set, thus computing 
the Casimir energy of a simple fractal for the first time.

A fractal often has unusual scaling behavior, which often leads to
noninteger fractal dimensions for volume, area, or perimeter for
these geometric shapes. The Casimir energy also depends on the geometry
of the cavity that binds the field. In this context,
the connections between the Casimir energy and the Weyl's problem
on the asymptotic distribution of the eigenvalues for the wave equation
for smooth boundaries is well documented, for example, see 
Refs.~\cite{Baltes1978sf,Balian1970401,Balian1971271}.
Berry in Refs.~\cite{Berry1979ps,Berry1996bf} 
conjectured that the Weyl's formula
in Ref.~\cite{weyl} for the asymptotic mode number extends 
for fractal regions and/or surfaces.
This Berry-Weyl conjecture has been shown to hold, 
if the dimensions of the regions and surfaces are interpreted as 
the Minkowski-Bouligand dimension~\cite{Lapidus1988lp}
instead of the Hausdorff-Besicovitch dimension as originally proposed by Berry.
The example consisting of parallel plates positioned at the points
of a Cantor set has the dimension for its boundary equal to 2 because
it is bounded by two-dimensional planes, and the volume dimension
of the Cantor set is $2+\ln 2/\ln 3\sim 2.63093$.
The suggestion seems to be that it might be possible to read out the 
fractal dimension of a region from its Casimir energy~\cite{Kigami1993fs}.
In the example of the Cantor set, the total single-body energy is given by
$\sum_{i}\Delta {\cal E}_i$.
For identical plates, $\Delta {\cal E}_i$ is the same for all the plates.
Thus, it factors out of the sum, and the remaining sum involves 
the addition of all the points of the Cantor set,
which is suggestive evidence of the Weyl-Berry conjecture. 
 
The only Casimir energy calculation that has been achieved for 
an infinite stack of plates before our work is probably that of
equidistant parallel plates, in the spirit of our discussion in 
Sec.\,\ref{sec-rel-string}. Using periodic boundary conditions,
dictated by the periodicity of the plates, the problem reduces to finding 
the dispersion relation that determines the modes.
Having described a formalism that could be used to work with configurations
that does not involve equidistant plates, one could now entertain the
idea of calculating the Casimir energy of a quasi-crystal. 
The remarks on the Poisson summation formula in the context of 
a quasi-crystal in Ref.~\cite{Ninham:ab0253}
and on temperature inversion symmetry in the finite-temperature Casimir effect
in Ref.~\cite{Ravndal1989ec} might be indicative of this possibility.

\appendix*
\section{Proof of the Faddeev Equations (\ref{tm-sol})}
\label{sec-Faddeev-Eqn-proof}

Operating two derivatives with respect to $z$ in the 
ansatz of Eq.~(\ref{ans-Ngf}), we obtain
\begin{equation}
\frac{d^2}{dz^2} g_{1\ldots N}(z,z^\prime) = \frac{d^2}{dz^2} g_0(z-z^\prime)
-\frac{d^2}{dz^2} {\bf r}(z) \cdot {\bf t}_{1\ldots N} \cdot {\bf r}(z^\prime),
\end{equation} 
which using the differential equations for 
$g_{1\ldots N}(z,z^\prime)$, $g_0(z-z^\prime)$, and $r_i(z)$,
in Eqs.~(\ref{gf-Ndplates}), (\ref{gf-free}), and (\ref{ri-def}),
leads to the relation
\begin{equation}
\sum_{i=1}^N \lambda_i \,\delta(z-a_i) g_{1\ldots N}(z,z^\prime)
= \sum_{i=1}^N \sum_{j=1}^N \delta(z-a_i)\, t_{ij}\, g_0(z^\prime-a_j).
\label{ldg=dtg}
\end{equation}
Integrating Eq.~(\ref{ldg=dtg}) over $z$ from $z=a_i-\delta$ to $z=a_i+\delta$
for small $\delta$, we have
\begin{equation}
\lambda_i \, g_{1\ldots N}(a_i,z^\prime) 
= \sum_{j=1}^N t_{ij}\, g_0(z^\prime-a_j)
\label{lgn=tg0}
\end{equation}
in which there is no summation on $i$. At this point, we note that 
these Green's functions satisfy the reciprocity theorem
\begin{equation}
g_{1\ldots N}(z,z^\prime) = g_{1\ldots N}(z^\prime,z),
\end{equation}
which requires the transition matrix to be symmetric,
\begin{equation}
t_{ij}=t_{ji}.
\end{equation}
We, of course, also have
\begin{equation}
g_0(z-z^\prime) = g_0(z^\prime-z).
\end{equation}
We use the ansatz in Eq.~(\ref{ans-Ngf}) to replace the left-hand side
of Eq.~(\ref{lgn=tg0}), operate it with two derivatives with respect
to $z^\prime$, and use the differential equation for $g_0(z-z^\prime)$
in Eq.~(\ref{gf-free}) in conjunction with the reciprocal symmetry of
Green's function to derive
\begin{eqnarray}
\lambda_i \left[ \delta(a_i-z^\prime) 
- \sum_{m=1}^N \sum_{j=1}^N g_0(a_i-a_m)\, t_{mj}\, \delta(a_j-z^\prime)\right]
\nonumber \\ =\sum_{j=1}^N t_{ij}\, \delta(z^\prime-a_j). \hspace{15mm}
\label{ld-gtd=td}
\end{eqnarray}
Integrating Eq.~(\ref{ld-gtd=td}) over $z^\prime$ 
from $z^\prime=a_j-\delta$ to $z^\prime=a_j+\delta$ for small $\delta$, we have
\begin{equation}
\lambda_i \left[ \delta_{ij} 
- \sum_{m=1}^N g_0(a_i-a_m) \, t_{mj} \right] = t_{ij},
\end{equation}
which when rearranged, using Eq.~(\ref{dimless-def}),
and expressed in vector notation is the Faddeev equation in Eq.~(\ref{tm-sol}).

It is also instructive to express the Faddeev equation in
Eq.~(\ref{tm-sol}) in terms of single-plate transition matrices.
In terms of the diagonal matrix
\begin{equation}
{\bf t}_\text{diag} = \left[ \begin{array}{cccc}
t_1 & ~ &  & ~ \\
 &t_2  & \rput(0.3,0.2){\text{\huge 0}} &  \\
 &  & \ddots & \\
\rput(0.2,0.4){\text{\huge 0}} &  & & t_N
\end{array} \right]
\end{equation}
the matrix equation of Eq.~(\ref{tm-sol}) can be rewritten in the form 
\begin{equation}
\tilde {\bf t}_{1\ldots N} = \left[ {\bf 1}
+ \tilde {\bf t}_\text{diag} \cdot (\tilde{\bf R} -{\bf 1}) \right]^{-1}
\cdot \tilde {\bf t}_\text{diag},
\label{tm-solt}
\end{equation}
which is more easily solved.

\acknowledgments

We dedicate this work to the memory of Martin Schaden, who passed away
while this paper was being refereed. The ideas in the present paper emerged
from Martin's work in Refs.\cite{Schaden:2010wv,Shajesh:2011ef}, 
and we remember him for the collaborative assistance.
K.V.S. would like to thank Jerzy Kocik and P. Sivakumar for discussions 
on the Apollonian gasket, which directly led to conducting this study.
We thank Mathias Bostr\"{o}m and Jose M. Mu\~{n}oz-Casta\~{n}eda
for feedback on the manuscript and pointing us to relevant references.
We acknowledge support from the Research Council of Norway (Project No. 250346).
I.C.P. acknowledges support from Centro Universitario de la Defensa
(Grant No. CUD2015-12), Spanish MINECO/FEDER (Grant  No. FPA2015-65745-P),
and DGA-FSE (Grant No. 2015-E24/2).

\bibliography{biblio/b20150531-self-similar-plates,%
biblio/b20150531-self-similar-plates-Iver}

\end{document}